\documentclass{PoS}

\title{New results from the lattice on the theoretical inputs to
the hadronic $\tau$ determination of $V_{us}$}

\ShortTitle{Lattice input on the $\tau$ decay determination of $V_{us}$}

\author{P.A. Boyle,$^a$, L. Del Debbio,$^a$ N. Garron,$^b$,
R.J. Hudspith,$^a$,
E. Kerrane$^c$, \speaker{K. Maltman},$^{d,e}$ and J.M. Zanotti$^e$
\\
\llap{$^a$}Physics and Astronomy, University of Edinburgh, Edinburgh
EH9 3JZ, UK\\
\llap{$^b$}School of Mathematics, Trinity College, Dublin 2, Ireland\\
\llap{$^c$}Instituto de F\`isica T\`eorica UAM/CSIC, Universidad
Aut\`onoma de Madrid, Cantoblanco E-28049 Madrid, Spain\\
\llap{$^d$}Mathematics and Statistics, York University, Toronto M3J 1P3
Canada\\
\llap{$^e$}CSSM, University of Adelaide, Adelaide 5005 Australia\\
E-mail: \email{paboyle@ph.ed.ac.uk},
\email{ldeldebb@ph.ed.ac.uk}, \email{ngarron@maths.tcd.ie},
\email{s0968574@sms.ed.ac.uk}, \email{eoin.kerrane@gmail.com},
\email{kmaltman@yorku.ca}, \email{james.zanotti@adelaide.edu.au}}

\abstract{Recent sum rule determinations of $\vert V_{us}\vert$, 
employing flavor-breaking combinations of hadronic $\tau$ decay data,
are significantly lower than either expectations based on 3-family unitarity 
or determinations from $K_{\ell 3}$ and $\Gamma [K_{\mu 2}]
/\Gamma [\pi_{\mu 2}]$. We use lattice data to investigate 
the accuracy/reliability of the OPE representation of the 
flavor-breaking correlator combination entering the $\tau$ decay analyses. 
The behavior of an alternate correlator combination,
constructed to reduce problems associated with the slow
convergence of the $D=2$ OPE series, and entering an alternate sum
rule requiring both electroproduction cross-section and hadronic
$\tau$ decay data, is also investigated. Preliminary
updates of both analyses, with the lessons learned from the lattice
data in mind, are also presented.}

\FullConference{Xth Quark Confinement and the Hadron Spectrum,\\
		October 8-12, 2012\\
		TUM Campus Garching, Munich, Germany}

\begin{document}
\section{Background}
\label{intro}

The determination of $\vert V_{us}\vert$ from analyses of flavor-breaking (FB)
combinations of hadronic $\tau$ decay data~\cite{gamizetalvus,kmcwvus} 
proceeds via finite energy sum rules (FESRs), generically
\begin{equation}
\int_0^{s_0}w(s) \rho(s)\, ds\, =\, -{\frac{1}{2\pi i}}\oint_{\vert
s\vert =s_0}w(s) \Pi (s)\, ds\ ,
\label{basicfesr}
\end{equation}
the $\vert V_{us}\vert$ determination involving the FB 
difference $\Delta\Pi_\tau \, \equiv\,
\left[ \Pi_{V+A;ud}^{(0+1)}\, -\, \Pi_{V+A;us}^{(0+1)}\right]$,
with $\Pi^{(J)}_{V/A;ij}(s)$ the spin $J=0,1$ components
of the flavor $ij$, vector (V) or axial vector (A) current-current 2-point
functions. The spectral functions, $\rho^{(J)}_{V/A;ij}$,
of $\Pi^{(J)}_{V/A;ij}(s)$, and hence that, $\Delta\rho_\tau$, 
of $\Delta\Pi_\tau$, are related to the normalized differential distributions,
$dR_{V/A;ij}/ds$, of flavor $ij$ V- or
A-current-induced $\tau$ decay widths,
$R_{V/A;ij}\, \equiv\, \Gamma [\tau^- \rightarrow \nu_\tau
\, {\rm hadrons}_{V/A;ij}\, (\gamma )]/ \Gamma [\tau^- \rightarrow
\nu_\tau e^- {\bar \nu}_e (\gamma)]$, by~\cite{tsai}
\begin{eqnarray}
&&{\frac{dR_{V/A;ij}}{ds}}\, =\, c^{EW}_\tau \vert V_{ij}\vert^2
\left[ w_\tau (s ) \rho_{V/A;ij}^{(0+1)}(s)
- w_L (s )\rho_{V/A;ij}^{(0)}(s) \right]
\label{basictaudecay}\end{eqnarray}
with $w_\tau (s)$, $w_{L}(s)$ and $c^{EW}_\tau$ all known,
and $V_{ij}$ the flavor $ij$ CKM matrix element.
With $\vert V_{ud}\vert$ from other sources,
$\Delta\rho_\tau (s)$ is expressible in terms of
experimental data and $\vert V_{us}\vert$. $\vert V_{us}\vert$
is then obtained by using the OPE for $\Delta\Pi_\tau$ on the RHS
and data on the LHS of Eq.~(\ref{basicfesr}).

The use of FESRs involving the $J=0+1$ combination $\Delta\Pi_\tau$
is necessitated by the very bad behavior of the integrated
$J=0$, $D=2$ OPE series at scales kinematically
accessible in $\tau$ decay~\cite{longprob}. Fortunately,
the dominant such $dR_{ud,us}/ds$ contributions are
from the $\pi$ and $K$ poles, whose strengths are 
accurately known. The remaining $J=0$ contributions, which are doubly chirally 
suppressed, are obtainable phenomenologically~\cite{jop,mksps}.
With $J=0$ contributions subtracted from $dR_{ud,us}/ds$,
one obtains $\rho_{V/A;ud,us}^{(0+1)}(s)$,
allowing the LHS of Eq.~(\ref{basicfesr}) to be formed for any
$w(s)$ and $s_0$. Defining the re-weighted $J=0+1$
spectral integrals $R^w_{V+A;ij}(s_0)=\int_0^{s_0}ds\, w(s)
dR^{(0+1)}_{V+A;ij}(s)/ds$,
\begin{equation}
\vert V_{us}\vert \, =\, \sqrt{R^w_{V+A;us}(s_0)/\left[
{\frac{R^w_{V+A;ud}(s_0)}{\vert V_{ud}\vert^2}}
\, -\, \delta R^{w,OPE}_{V+A}(s_0)\right]}\ ,
\label{tauvussolution}\end{equation}
where $ \delta R^w_{V+A}(s_0)\, =\,
{\frac{R^w_{V+A;ud}(s_0)}{\vert V_{ud}\vert^2}}
\, -\, {\frac{R^w_{V+A;us}(s_0)}{\vert V_{us}\vert^2}}$.
$\vert V_{us}\vert$ should be independent
of $w(s)$ and $s_0$, providing tests of the reliability of
the OPE treatment and input data employed. Recent
determinations~\cite{kmetcvustau}, which yield
$\vert V_{us}\vert$ $\sim 3\sigma$ lower than
3-family-unitarity expectations{\footnote{A recent update of the kinematic
weight, $s_0=m_\tau^2$ analysis, e.g., quotes the result
$\vert V_{us}\vert = 0.2173(20)_{exp}(10)_{th}$~\cite{gamizckm12}.}},
show non-trivial $w(s)$- and $s_0$-dependence, suggesting 
shortcomings in the experimental data and/or OPE representation.

Quantifying the OPE uncertainty and, from this,
the theoretical error on $\vert V_{us}\vert$, is complicated by the
slow convergence, at the correlator level, of the leading $D=2$
OPE series $\left[\Delta \Pi_\tau \right]_{D=2}^{OPE}$.
To four loops, with $\bar{a}=\alpha_s(Q^2)/\pi$, and
$\alpha_s(Q^2)$, $m_s(Q^2)$ the running coupling and strange quark
mass in the $\overline{MS}$ scheme, and neglecting $m_{u,d}$ relative
to $m_s$, one has, from Ref.~\cite{bckd2ope}{\footnote{We use the
estimate $d_4= 2378$ of Ref.~\cite{bckd2ope} for the at-present-unknown
5-loop coefficient $d_4$.}}
\begin{eqnarray}
&&\left[\Delta\Pi_\tau (Q^2)\right]^{OPE}_{D=2}\, =\, {\frac{3}{2\pi^2}}\,
{\frac{m_s(Q^2)}{Q^2}} \left[ 1\, +\, {\frac{7}{3}} \bar{a}\,
+\, 19.93 \bar{a}\,^2 \, +\, 208.75 \bar{a}\,^3
\, +\, d_4 \bar{a}\,^4\, +\, \cdots \right]\ .\ \
\label{d2form}\end{eqnarray}
Since $\bar{a}(m_\tau^2)\simeq 0.1$, convergence at the spacelike point on
the contour $\vert s\vert = s_0$ is marginal at best, and conventional error 
estimates may significantly underestimate the $D=2$ truncation uncertainty.
The alternate fixed-order (FOPT) and contour-improved (CIPT) schemes
for the truncated integrated series{\footnote{In FOPT,
one first integrates with fixed renormalization
scale $\mu$, then resums logs through the ``fixed-scale'' choice $\mu^2=s_0$;
in CIPT logs are instead resummed point by point along the contour before
integration via the ``local-scale'' choice $\mu^2 =Q^2$.}}, e.g.,
despite differing only by contributions beyond the
common truncation order, yield
$\vert V_{us}\vert$ whose difference not only significantly 
exceeds such estimates, but increases steadily (from
$\sim 0.0010$ to $\sim 0.0020$) as one moves from 3- to 5-loop
truncation.

With problems in the FB $\Delta\Pi_\tau$ FESRs due,
to at least some extent, to slow $D=2$ OPE convergence, 
FESRs having reduced $D=2$ OPE contributions 
at the correlator level are highly desirable.
In Ref.~\cite{kmtauem08}, FB combinations of $\Pi^{(0+1)}_{V/A;ud}$,
$\Pi^{(0+1)}_{V+A;us}$ and the EM correlator, $\Pi_{EM}$,
(whose spectral function, $\rho_{EM}$, is determined
by the bare $e^+e^-\rightarrow hadrons$ cross-sections)
were constructed having vanishing leading
$O(\alpha_s^0)$ $D=2$ OPE contributions.
The unique such combination in which $\Pi^{(0+1)}_{V+A;us}$
appears with the same normalization as in $\Delta\Pi_\tau$ is
\begin{equation}
\Delta\Pi_{\tau ,EM}= 9\Pi_{EM} -6\Pi^{(0+1)}_{V;ud} + \Delta\Pi_\tau
\end{equation}
whose $D=2$ OPE series is
\begin{equation}
{\frac{-3}{2\pi^2}}\,
{\frac{\bar{m}_s}{Q^2}} \left[ {\frac{1}{3}} \bar{a}
+4.38\bar{a}\,^2  + 44.9 \bar{a}\,^3+\cdots\right]\ .
\label{tauemd2form}\end{equation}
The higher order coefficients in this series
are also significantly smaller than those
for $\Delta\Pi_\tau$. The $D=4$ series is also, fortuitously,
suppressed, the results of Ref.~\cite{bnp} leading to the form
\begin{eqnarray}
&&{\frac{m_s\langle \bar{s}s\rangle\,
-\, m_\ell \langle \bar{\ell}\ell\rangle}{Q^4}}\sum_k c_k\bar{a}^k\ ,
\label{d4form}\end{eqnarray}
with $(c_0, c_1, c_2)=(-2,-2,-26/3)$ for $\Delta\Pi_\tau$
and $(0, 8/3, 59/3)$ for $\Delta\Pi_{\tau ,EM}$.
The analogue of Eq.~(\ref{tauvussolution})
for $\vert V_{us}\vert$, based on the $\Delta\Pi_{\tau ,EM}$ rather than
$\Delta\Pi_\tau$ FESR, is thus expected
to have a much smaller OPE contribution, and hence
much reduced theoretical uncertainty. Lattice data will be used to check
whether or not this expected OPE suppression is realized below.

\begin{figure}[!ht]
\includegraphics[width=0.75\columnwidth,angle=270]
{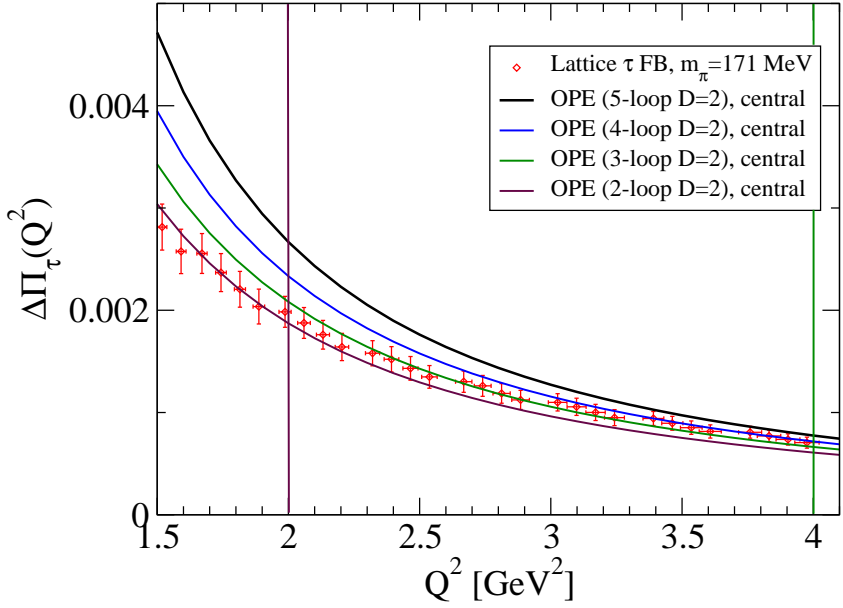}
\vspace{-0.95cm}
\caption{Lattice data vs. the OPE for $\Delta\Pi_\tau (Q^2)$}
\label{taufblattvsope}
\end{figure}

\section{Lattice vs OPE results for $\Delta\Pi_\tau$ and
$\Delta\Pi_{\tau ,EM}$}
The $\Pi_{V/A;ud}^{(J)}(Q^2)$ for spacelike $Q^2\, =\, -q^2>0$ also
enter the decomposition of the Euclidean space V and A 2-point functions,
and hence are measurable on the lattice.
We report here on comparisons of OPE expectations
for the combinations $\Delta\Pi_\tau (Q^2)$ and $\Delta\Pi_{\tau ,EM}(Q^2)$ with
results obtained on $n_f=2+1$ domain wall fermion
ensembles with $1/a=1.37\ GeV$ and
$m_\pi = 171$ and $248\ MeV$. Full details
of the simulations are given in Ref.~\cite{ainv137}.
Expanded comparisons to results from finer $1/a=2.28\ GeV$, 
$m_\pi = 289,\, 345$ and $394\ MeV$ ensembles~\cite{ainv228},
will be considered elsewhere. Here, by
keeping momentum components $\leq$ $1/8^{th}$ of the lattice
maximum, $Q^2\sim 4.6\ GeV^2$ can be reached.
The OPE-lattice comparisons are designed to explore
(i) the accuracy of the OPE representation for different $D=2$ truncation
orders, (ii) the question of whether the fixed-scale or local-scale
representation best describes the $Q^2$-dependence of the lattice data,
and (iii) whether the data bears out the strong suppression of
$\Delta\Pi_{\tau ,EM}$ relative to $\Delta\Pi_\tau$ suggested by the
truncated OPE representations. Since results for $m_\pi =171,\, 248\ MeV$
are qualitatively identical, we show results for the former only.

We begin with the lattice-OPE comparison for the case where the 
dominant $D=2$ OPE contribution is evaluated using the local-scale prescription
(the analogue of the FESR CIPT prescription, used in essentially all
FESR $\vert V_{us}\vert$ determinations in the literature). 
Fig.~\ref{taufblattvsope} shows the OPE results for
2-, 3-, 4- and 5-loop $D=2$ truncation. An apparently
asymptotic behavior is found for the integrated $D=2$ OPE series,
the terms decreasing in magnitude with increasing order until the smallest 
term is reached, and increasing in magnitude thereafter. 
To the right of the right-most vertical line, the 4-loop 
contribution is smallest, and, interpreting the
behavior as that of a conventional asymptotic series, 
4-loop truncation would be favored. Between the two vertical lines 
the 3-loop contribution is smallest, favoring 3-loop truncation, 
while to the left of the left-most vertical line, 
where it is the 2-loop contribution which is smallest,
2-loop truncation is favored.

\begin{figure}[!ht]
\includegraphics[width=0.75\columnwidth,angle=270]
{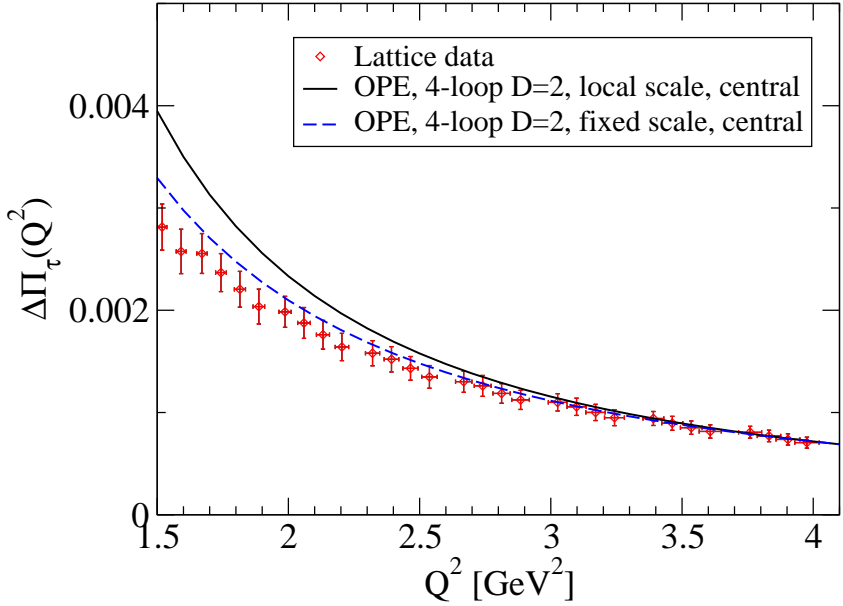}
\vspace{-0.95cm}
\caption{Lattice data vs. the OPE with
either fixed-scale or local-scale versions of the $D=2$
contribution to  $\Delta\Pi_\tau (Q^2)$}
\label{taufbfixedvslocalscaleope}
\end{figure}

In Fig.~\ref{taufbfixedvslocalscaleope}, we compare the $Q^2$-dependences
of the lattice data and OPE representation, the $D=2$
contribution to the latter being truncated at 4-loops and evaluated
using both local-scale and fixed-scale prescriptions.
For the fixed-scale case, we use $\mu^2=4\ GeV^2$.
The lattice data is evidently represented considerably better
by the fixed-scale version (whose use generates the FOPT
version of the FESR integrals).

\begin{figure}[!ht]
\includegraphics[width=0.75\columnwidth,angle=270]
{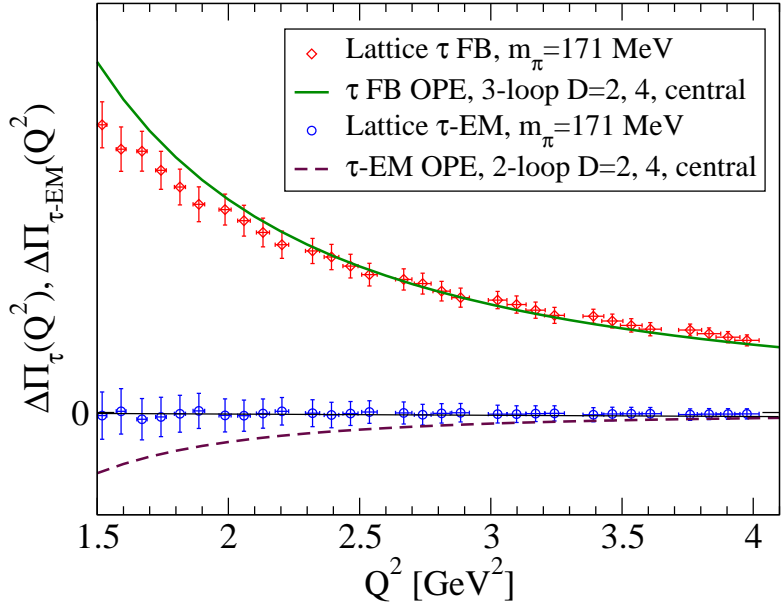}
\vspace{-0.95cm}
\caption{Lattice data vs. the OPE for $\Delta\Pi_\tau (Q^2)$, $\Delta\Pi_{\tau
    \
,EM}(Q^2)$}
\label{taufbtauemlattvsope}
\end{figure}

Fig.~\ref{taufbtauemlattvsope} compares the lattice data and OPE
representation for $\Delta\Pi_{\tau ,EM}(Q^2)$, with the analogous
$\Delta\Pi_\tau (Q^2)$ results included for comparison. The lattice
data clearly confirms the strong suppression in 
$\Delta\Pi_{\tau ,EM}$ relative to $\Delta\Pi_\tau$ suggested by
the OPE representation, the numerical extent of the suppression being
even greater than suggested by the central OPE result.
One thus expects very small theoretical errors on the $\vert V_{us}\vert$ 
obtained from mixed $\tau$-electroproduction FESRs.

We now perform preliminary updates of the $\Delta\Pi_\tau$ and
$\Delta\Pi_{\tau ,EM}$ FESR determinations of $\vert V_{us}\vert$,
taking the lessons provided by the above comparisons into account.
The flavor $us$ V+A $\tau$ data used are obtained by rescaling
the old ALEPH~\cite{alephus} results mode-by-mode for subsequent 
changes in the exclusive branching 
fractions. The updated branching fractions 
are those from the unitarity-constrained HFAG fit
incorporating also $K_{\mu 2}$ and $\pi_{\mu 2}$ input, 
discussed in Ref.~\cite{dv712},
further updated for the $B[\tau\rightarrow K^- n\pi^0\nu_\tau ]$
results of Ref.~\cite{adametz}. For the flavor $ud$ V and A
distributions, we employ the update of the OPAL distribution~\cite{opalud}
detailed in Ref.~\cite{dv712}, further modified by a small
common global V and A rescaling, needed to restore unitarity after
inclusion of the new $B[\tau\rightarrow K^- n\pi^0\nu_\tau ]$ results.
This interim global rescaling will be replaced
by a further-updated mode-by-mode rescaling once
the results of Ref.~\cite{adametz} are finalized and can be
incorporated into the global HFAG branching function fit.
While details of the electroproduction cross-section data employed
will be given elsewhere, we mention 
that the tension between $\tau$ and electroproduction
results for the $\pi\pi$ contribution to $\rho_{EM}^{I=1}(s)$
is assumed to be accounted for by the long-distance
EM $\rho -\gamma$ mixing effect identified in Ref.~\cite{jsrhogamma},
implying that the $\tau$ $\pi\pi$ data is to be used for
the $\pi\pi$ contribution to the
$\Delta\Pi_{\tau ,EM}$ FESR, where the effect of this long-distance EM
contribution is not accounted for on the OPE side.

\begin{figure}[!ht]
\includegraphics[width=0.75\columnwidth,angle=270]
{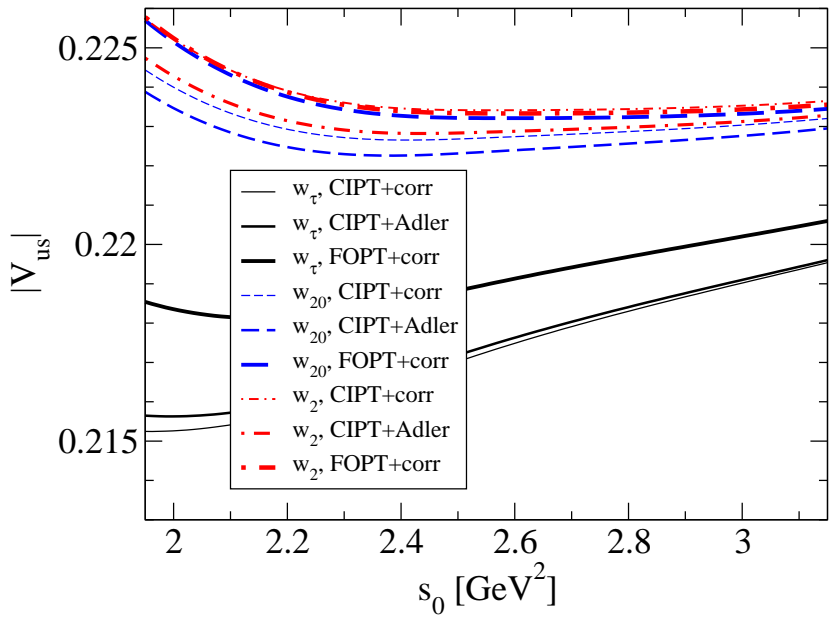}
\vspace{-0.95cm}
\caption{$V_{us}$ vs. $s_0$ for the $\tau$ FB sum rule}
\label{taufbsrvus}
\end{figure}

For the $\Delta\Pi_\tau$ analysis, we employ the 3-loop $D=2$ truncation
favored by the lattice data. The results obtained from FESRs involving
the kinematic weight, $w_\tau$, and two other weights 
used previously in the literature{\footnote{$w_{20}$ was constructed to
improve integrated $D=2$ CIPT convergence~\cite{kmjk00};
$w_2$ is a member of the family
$w_N(y)=1-{\frac{N}{N-1}}y+{\frac{1}{N-1}}y^{N}$
constructed to keep higher $D$ OPE contributions under
control~\cite{kmtyalphas}.}} are shown in Fig.~\ref{taufbsrvus},
for both CIPT and FOPT prescriptions, though it is the latter which is, 
in fact, favored. For CIPT, we show results obtained using
both the integrated correlator and same-order-truncated Adler function 
forms (the latter obtained by partial
integration and re-truncation before integration). These
again differ only by contributions beyond the common truncation order. 
For $s_0=m_\tau^2$, $w=w_\tau$, shifting from the 5-loop-truncated 
$D=2$ CIPT+correlator to the 3-loop-truncated FOPT prescription favored
by the lattice data raises $\vert V_{us}\vert$ by $0.0017$.
Significant $w(s)$-dependence, and, for $w_\tau$,
significant $s_0$-dependence, are evidently present,
though the latter is reduced when FOPT, rather than CIPT, is used
for the integrated $D=2$ series. 
These effects produce a contribution to the theoretical systematic uncertainty 
on $\vert V_{us}\vert$ already much larger than the 
total estimated theoretical uncertainty 
reported previously in the literature.

\begin{figure}[!ht]
\includegraphics[width=0.75\columnwidth,angle=270]
{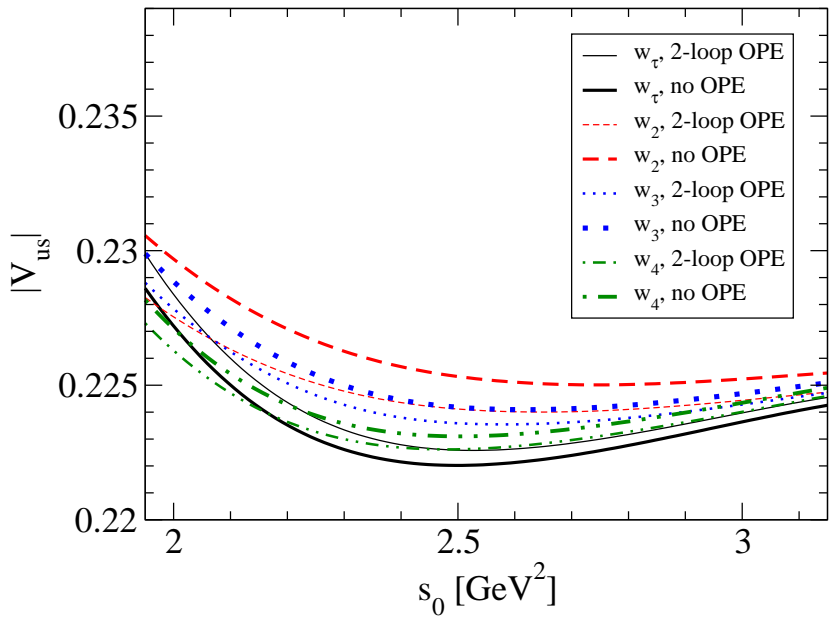}
\vspace{-0.95cm}
\caption{$V_{us}$ vs. $s_0$ for the mixed $\tau$-EM sum rule}
\label{tauemsrvus}
\end{figure}

Fig.~\ref{tauemsrvus} shows the results for $\vert V_{us}\vert$
obtained from $\Delta\Pi_{\tau EM}$ FESRs employing a number of weights
used in the earlier literature~\cite{kmjk00,kmtyalphas}. 
There are two curves for each weight, one corresponding to an analysis 
in which (in keeping with the lattice results) OPE contributions are 
treated as negligible, one to an analysis using the 2-loop-truncated 
version of the $D=2$ OPE series. The results show much weaker 
$s_0$- and $w(s)$-dependence, and are in excellent agreement with
the expectations of 3-family unitarity. We emphasize that these results 
are preliminary, and require further updating once improved versions of 
the input experimental data become available.

\vskip .1in\noindent
{\bf Acknowledgements:} The computations were done using the STFC's DiRAC
facilities at Swansea and Edinburgh. PAB, LDD, NG and RJH are supported
by an STFC Consolidated Grant, and by the EU under Grant Agreement
PITN-GA-2009-238353 (ITN STRONGnet). EK was supported by the Comunidad
Aut\`onoma de Madrid under the program HEPHACOS S2009/ESP-1473 and the
EU under Grant Agreement PITN-GA-2009-238353 (ITN STRONGnet).
KM acknowledges the hospitality of the CSSM, University of Adelaide,
and support of NSERC (Canada).
JMZ is supported by the Australian Research Council grant FT100100005.


\begin{thebibliography}{99}
\bibitem{gamizetalvus}E. Gamiz {\it et al.}, JHEP 0301 (2003) 060;
Phys. Rev. Lett. 94 (2005) 011803.
\bibitem{kmcwvus}K. Maltman and C.E. Wolfe, Phys. Lett. B639
(2006) 283; {\it ibid.} B650 (2007) 27; K. Maltman {\it et al.}
Int. J. Mod. Phys. A23 (2008) 3191.
\bibitem{tsai}Y.-S. Tsai, Phys. Rev. D4 (1971) 2821.
\bibitem{longprob}K. Maltman, Phys. Rev. D58 (1998) 093015;
K. Maltman and J. Kambor, {\it ibid.} D64 (2001) 093014.
\bibitem{jop}M. Jamin, J.A. Oller and A. Pich, Nucl. Phys. B587
(2000) 331; {\it ibid.} B622 (2002) 279; and Phys. Rev.
D74 (2006) 074009.
\bibitem{mksps}K. Maltman and J. Kambor, Phys. Rev. D65 (2002)
074013.
\bibitem{gamizckm12}E. Gamiz, talk presented at CKM 2012.
\bibitem{kmetcvustau}K. Maltman, {\it et al.}, Nucl. Phys. Proc. Suppl. 189
(2009) 175; K. Maltman, {\it ibid.} 218 (2011) 146.
\bibitem{bckd2ope}K.G. Chetyrkin and A. Kwiatkowski, Z. Phys. C59 (1993) 525
and hep-ph/9805232; P.A. Baikov, K.G. Chetyrkin and
J.H. Kuhn, Phys. Rev. Lett. 95 (2005) 012003.
\bibitem{bnp}E. Braaten, S. Narison, A. Pich, {\it Nucl. Phys.} {\bf B373}
(1992) 581.
\bibitem{kmtauem08}K. Maltman, Phys. Lett. B672 (2009) 257.
\bibitem{ainv137}R. Arthur {\it et al.}, arXiv:1208.4412 (hep-lat).
\bibitem{ainv228}Y. Aoki {\it et al.}, \emph{Phys. Rev.} {\bf D83} (2011)
074508 [{\tt arXiv:1011.0892 (hep-lat)}].
\bibitem{alephus}R. Barate {\it et al.} (ALEPH Collaboration),
\emph{Eur. Phys. J.} {\bf C11} (1999) 599.
\bibitem{dv712}D. Boito {\it et al.}, \emph{Phys. Rev.} {bf D85} (2012)
093015 [{\tt arXiv:1203.3146 (hep-ph)}].
\bibitem{adametz}A. Adametz, ``Measurement of $\tau$ decays into
charged hadron accompanied by neutral $\pi$ mesons and determination of
the CKM matrix element $\vert V_{us}\vert$'', University of Heidelberg
PhD thesis, July 2011 and the BaBar Collaboration, in progress
\bibitem{jsrhogamma}F. Jegerlehner and R. Szafron, \emph{Eur. Phys.
J.} {\bf C71} (2011) 1632 [{\tt arXiv:1101.2872 (hep-ph)}].
\bibitem{opalud}K. Ackerstaff {\it et al.} (OPAL Collaboration),
\emph{Eur. Phys. J.} {\bf C7} (1999) 271 [{\tt arXiv:hep-ex/9808019}].
\bibitem{kmjk00}J. Kambor and K. Maltman, \emph{Phys. Rev.} {\bf D62}
(2000) 093023 [{\tt hep-ph/0005156}].
\bibitem{kmtyalphas}K. Maltman and T. Yavin, \emph{Phys. Rev.} {\bf D78}
(2008) 094020 [{\tt arXiv:0807.0650}].


\end{thebibliography}
\end{document}